\documentclass[letter]{aa} 
\usepackage{graphicx}
\usepackage{txfonts}
\begin{document}
   \title{
B fields in OB stars (BOB): The discovery of a magnetic field in a multiple system in the Trifid Nebula,
one of the youngest star forming regions\thanks{Based
on observations obtained in the framework of the ESO Prg.\ 191.D-0255(A,B).}}
\titlerunning{The magnetic field of HD\,164492C}

   \author{
S.~Hubrig\inst{1}
\and L.~Fossati\inst{2}
\and T.~A.~Carroll\inst{1}
\and N.~Castro\inst{2}
\and J.~F.~Gonz\'alez\inst{3}
\and I.~Ilyin\inst{1}
\and N.~Przybilla\inst{4}
\and M.~Sch\"oller\inst{5}
\and L.~M.~Oskinova\inst{6}
\and T. Morel\inst{7}
\and N.~Langer\inst{2}
\and R.~D.~Scholz\inst{1}
\and N.~V.~Kharchenko\inst{8}
\and M.-F. Nieva \inst{4,9}
\and the BOB collaboration
}

\authorrunning{Hubrig et al.}

\institute{Leibniz-Institut f\"ur Astrophysik Potsdam (AIP), An der Sternwarte 16, 14482 Potsdam, Germany\\
              \email{shubrig@aip.de}
\and Argelander-Institut f\"ur Astronomie, Universit\"at Bonn, Auf dem H\"ugel~71, 53121~Bonn, Germany
\and Instituto de Ciencias Astronomicas, de la Tierra, y del Espacio (ICATE), 5400~San~Juan, Argentina
\and Institute for Astro- and Particle Physics, University of Innsbruck, Technikerstr.~25/8, 6020~Innsbruck, Austria
\and European Southern Observatory, Karl-Schwarzschild-Str.~2, 85748~Garching, Germany
\and Universit\"at Potsdam, Institut f\"ur Physik und Astronomie, 14476~Potsdam, Germany 
\and Institut d'Astrophysique et de G\'eophysique, Universit\'e de Li\`ege, All\'ee du~6~Ao\^ut, B\^at.~B5c, 4000~Li\`ege, Belgium
\and Main Astronomical Observatory, 27~Academica Zabolotnogo Str., 03680~Kiev, Ukraine
\and Dr.~Karl Remeis-Observatory \& ECAP, University Erlangen-Nuremberg, Sternwartstr.~7, D-96049 Bamberg, Germany
}
\date{Received; accepted}

  \abstract
   {}
   {
Recent magnetic field surveys in O- and B-type stars revealed that about
10\% of the core-hydrogen-burning massive stars host large-scale magnetic fields.
The physical origin of these fields is highly debated.
To identify and model the physical processes responsible for the generation of magnetic fields in massive
stars, it is important to establish whether magnetic massive stars are found in very young star-forming regions
or whether they are formed in close interacting binary systems.
}
   {
 In the framework of our ESO Large Program, we carried out low-resolution spectropolarimetric observations with 
FORS\,2 in 2013 April of the three most massive 
central stars in the Trifid nebula, HD\,164492A, HD\,164492C, and HD\,164492D. These observations indicated 
a strong longitudinal magnetic field of about 500--600\,G in the poorly studied component HD\,164492C. 
To confirm this detection,
we used HARPS in spectropolarimetric mode on two consecutive nights in 2013 June.
}
{
Our HARPS observations confirmed the longitudinal magnetic field in HD\,164492C. 
Furthermore, the HARPS observations revealed that HD\,164492C cannot be considered as a single star as it
possesses one or two companions. The 
spectral appearance indicates that the primary is most likely of spectral type B1--B1.5\,V.
Since in both observing nights most spectral lines appear blended,
it is currently unclear which components are magnetic. 
Long-term monitoring using high-resolution
spectropolarimetry is necessary to separate the contribution of each component to the magnetic signal.
Given the location of the system HD\,164492C in one of the youngest star formation regions,
this system can be considered as a Rosetta Stone for our understanding of the origin of magnetic fields in massive stars.
}
   {}


   \maketitle

\section{Introduction}

Magnetic fields have fundamental effects not only on the evolution of massive stars, on their rotation, 
and on the structure, dynamics, and heating of their radiatively-driven winds, but also on their final display as
supernova or gamma-ray burst.
About a few dozen massive magnetic stars are currently known, just enough to establish
the fraction of magnetic, core-hydrogen burning stars to be of the order of 8\% (Grunhut et al.\ \cite{grunhut2012}),
which appears to be similar to that of intermediate-mass stars.
While it is established that the magnetic fields in massive stars are
not dynamo-supported, but stable with decay times exceeding the stellar lifetime,
their origin is highly debated. The two main competing ideas
are that the fields are either ``fossil'' remnants of the Galactic ISM field that are amplified
during the collapse of a magnetised gas cloud (e.g.\ Price \& Bate\ \cite{price2007}), or that 
they are formed in a dramatic close-binary interaction,
i.e., in a merger of two stars or a dynamical mass transfer event (e.g.\ Ferrario et al.\ \cite{fer2009}).
The intermediate-mass stars show a magnetic fraction during their pre-main sequence evolution similar
to Herbig stars (e.g., Hubrig et al.\ \cite{Hubrig2009}; Hubrig et al.\ \cite{Hubrig2013} and references therein; Alecian et al.\ \cite{alecian13})
as their main-sequence descendants,
which may speak for fossil fields. On the other hand, there are almost no close binaries amongst the magnetic
intermediate-mass main-sequence Ap-type stars (e.g.\ Carrier et al.\ \cite{car2002}). Since this is expected if they were 
merger products, this argues for the binary hypothesis of the field origin. 

We present our spectropolarimetric observations of three massive stars in the Trifid Nebula in the framework of our 
``B fields in OB stars'' (BOB) collaboration.  This nebula is a very 
young ($\lesssim 10^6$\,yrs) and active site of 
star formation containing a rich population of young stellar objects (YSOs) and protostars 
(e.g.\ Cernicharo et al.\ \cite{cernicharo98}). The spectacular, well-known 
optical \ion{H}{ii} region provides an ideal place for investigating the onset of star birth 
and triggered star formation. This large nebula is ionised by the O7.5~Vz star HD\,164492A (Sota et al.\ \cite{Sota2013}),
which is the central object of the multiple system ADS~10991,
containing at least seven components (A to G; Kohoutek et al.\ \cite{koho99}).
Due to the faintness of the components HD\,164492B and HD\,164492E-G (all have visual magnitudes fainter than 10.6),
we only searched for a magnetic field for the three most massive components, HD\,164492A, C, and D.

\section{Magnetic field measurements using FORS\,2 spectropolarimetry}

\begin{figure}
\centering
\includegraphics[width=0.24\textwidth]{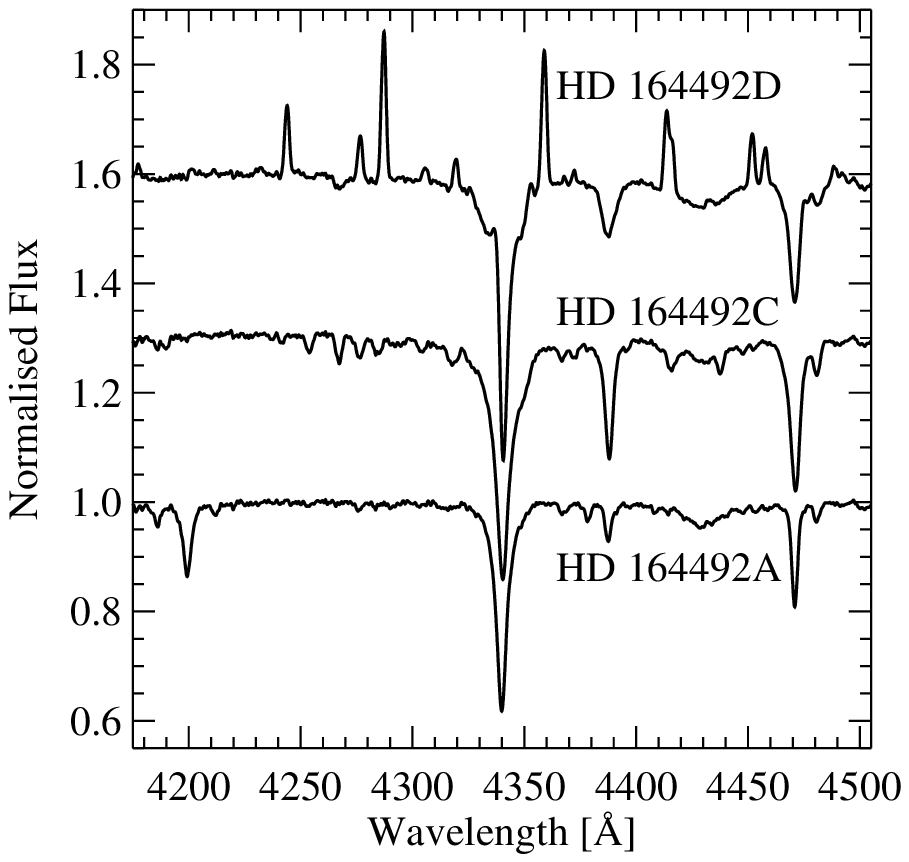}
\includegraphics[width=0.24\textwidth]{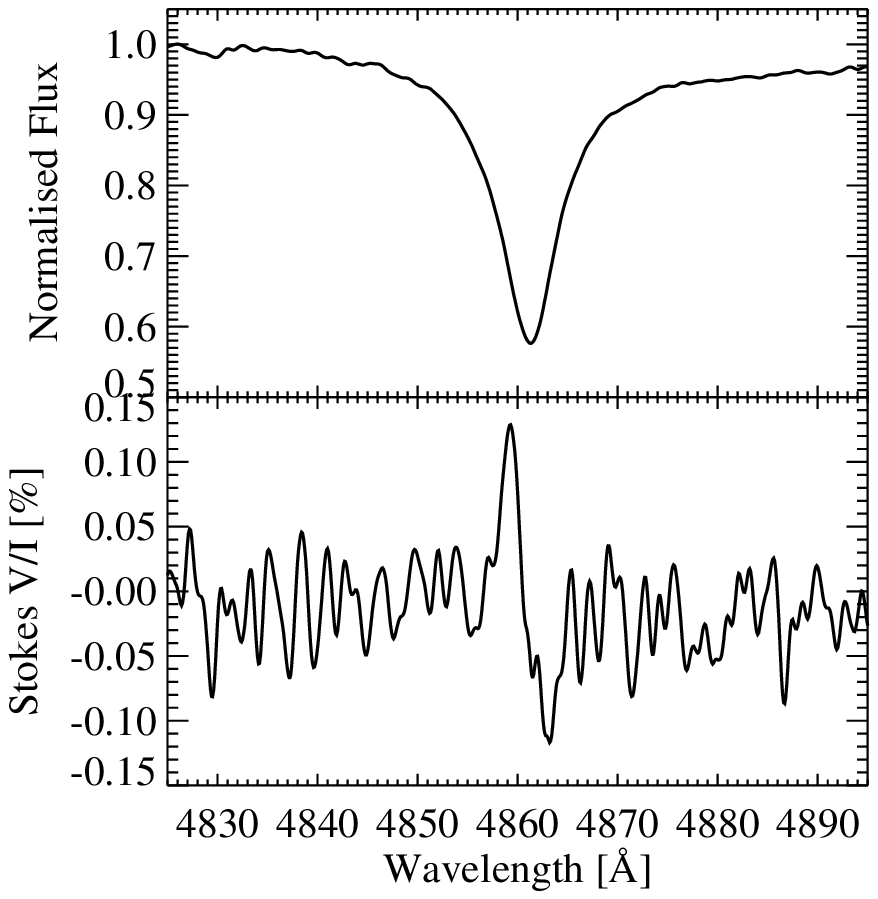}
\caption{
{\it Left panel}:
Stokes $I$ FORS\,2 spectra of three components of the multiple system ADS~10991 (HD\,164492A,  HD\,164492C, 
and HD\,164492D) in the vicinity of 
the H$\gamma$ line.
{\it Right panel}:
Stokes $I$ (top) and $V/I$ (bottom) FORS\,2 spectra of HD\,164492C in the vicinity of 
the H$\beta$ line.
}
\label{fig:fors}
\end{figure}

Three components of the system ADS~10991, HD\,164492A, C, and D, were observed with the
FOcal Reducer low dispersion Spectrograph (FORS\,2)
mounted on the 8\,m Antu telescope of the VLT
on 2013 April 9.
This multi-mode instrument is equipped with polarisation-analysing optics
comprising super-achromatic half-wave and quarter-wave phase retarder plates,
and a Wollaston prism with a beam divergence of 22$\arcsec$ in standard-resolution mode. 
We used the GRISM 600B and the narrowest available slit width
of 0$\farcs$4 to obtain a spectral resolving power of $R\sim2000$.
The observed spectral range from 3250 to 6215\,\AA{} includes all Balmer lines 
apart from H$\alpha$, and numerous \ion{He}{i} lines. The position angle of the retarder waveplate was changed from
$+45^{\circ}$ to $-45^{\circ}$ and vice versa every second exposure, i.e., we executed the
sequence $-45^{\circ}$$+45^{\circ}$, $+45^{\circ}$$-45^{\circ}$, $-45^{\circ}$$+45^{\circ}$, etc.\ up to 6--8 times.
For the observations we used a non-standard readout mode with low 
gain (200kHz,1$\times$1,low), which provides a broader dynamic range, hence 
allowed us to reach a higher signal-to-noise ratio (S/N) in the individual spectra.
We achieved a S/N of 1670 per pixel for the component HD\,164492A, 1490 for HD\,164492C, 
and 835 for HD\,164492D in the final integral spectra.
The integral spectra of all three components in the spectral region around H$\gamma$ are presented in the left panel of Fig.~\ref{fig:fors}. 
While the spectrum of component HD\,164492A corresponds to that of a typical O7-type star,
HD\,164492C appears to be an early-B type star,
and the spectrum of HD\,164492D displays numerous emission lines
similar to those observed in Herbig Be stars (e.g.\ Herbig \cite{herbig57}).

The mean longitudinal magnetic field is the average over the stellar hemisphere
visible at the time of observation of the component of the magnetic field
parallel to the line of sight, weighted by the local emerging spectral line
intensity.
The determination of the mean longitudinal magnetic field using low-resolution FORS\,1/2 spectropolarimetry 
has been described in detail by two different groups:  Bagnulo et al.\ (\cite{Bagnulo2002,Bagnulo2012}) and
Hubrig et al.\ (\cite{Hubrig2004a,Hubrig2004b}).
To identify systematic differences that might exist when the FORS\,2 data is treated by different groups,
the mean longitudinal 
magnetic field, $\left< B_{\rm z}\right>$, was derived in all three stars by each group separately, using 
independent reduction packages.
No magnetic field at a significance level of 3$\sigma$ was detected in HD\,164492A and HD\,164492D.
For HD\,164492C, using a set of IRAF and IDL routines based on the recipes described by 
Bagnulo et al.\ (\cite{Bagnulo2012}) and Fossati et al.\ (in prep.), 
we determined a mean longitudinal magnetic field of 
$\left<B_{\rm z}\right>_{\rm all}=523\pm37$\,G measured using the whole spectrum and
$\left<B_{\rm z}\right>_{\rm hyd}=600\pm54$\,G using only the hydrogen lines, i.e. the magnetic field is discovered
in this component at a significance level higher than 10$\sigma$. These measurements are consistent within a few tens 
of Gauss with those obtained using the software package developed by Hubrig et al.\ (\cite{Hubrig2004a,Hubrig2004b}): 
$\left<B_{\rm z}\right>_{\rm all}=472\pm44$\,G and
$\left<B_{\rm z}\right>_{\rm hyd}=576\pm60$\,G.
To illustrate the strong magnetic field in HD\,164492C, we present 
in the right panel of Fig.~\ref{fig:fors} the Stokes $I$ and $V$ spectra in which a distinct Zeeman 
feature is observed at the position of the H$\beta$ line.

\section{Magnetic field measurements using HARPS}

To further investigate the spectral appearance and behaviour of the magnetic field in HD\,164492C, 
we acquired two additional 
spectropolarimetric observations with the HARPS polarimeter (Snik et al.\ \cite{snik2008})
attached to ESO's 3.6\,m telescope (La Silla, Chile) 
on two consecutive nights at the beginning of 2013 June.
The polarimetric spectra with a $S/N$ of about 350 per pixel in the Stokes~$I$ spectra and a resolving 
power of $R = 115\,000$  
cover the spectral range  3780--6910\,\AA{}, with a small gap between 5259 and 5337\,\AA{}.
Each observation was split into four subexposures, 
obtained with different orientations of the quarter-wave retarder plate relative to the 
beam splitter of the circular polarimeter. Again, the reduction and magnetic field measurements 
were carried out using independent software packages developed for the treatment of HARPS data.

Within the first package,
the reduction and calibration was performed using the HARPS data reduction 
pipeline available at the 3.6\,m telescope in Chile. 
The normalisation of the spectra to the continuum level consisted of several steps described in detail 
by Hubrig et al.\ (\cite{Hubrig2013}).
The Stokes~$I$ and $V$ parameters were derived following the ratio method described by 
Donati et al.\ (\cite{Donati1997}), and null polarisation spectra 
were calculated by combining the subexposures 
in such a way that polarisation cancels out.
These steps ensure that the data contain no spurious signals (e.g.\ Ilyin \cite{Ilyin2012}).
Within the second software package, we reduced and calibrated the data with the {\sc REDUCE} 
package (Piskunov \& Valenti \cite{pisk02}), which performs an optimal extraction of the echelle orders after 
several standard steps, such as bias subtraction, flat-fielding, and cosmic-ray removal.
The wavelength calibration and the continuum normalisation were treated using
standard techniques.

\begin{figure}
\centering
\includegraphics[width=0.24\textwidth]{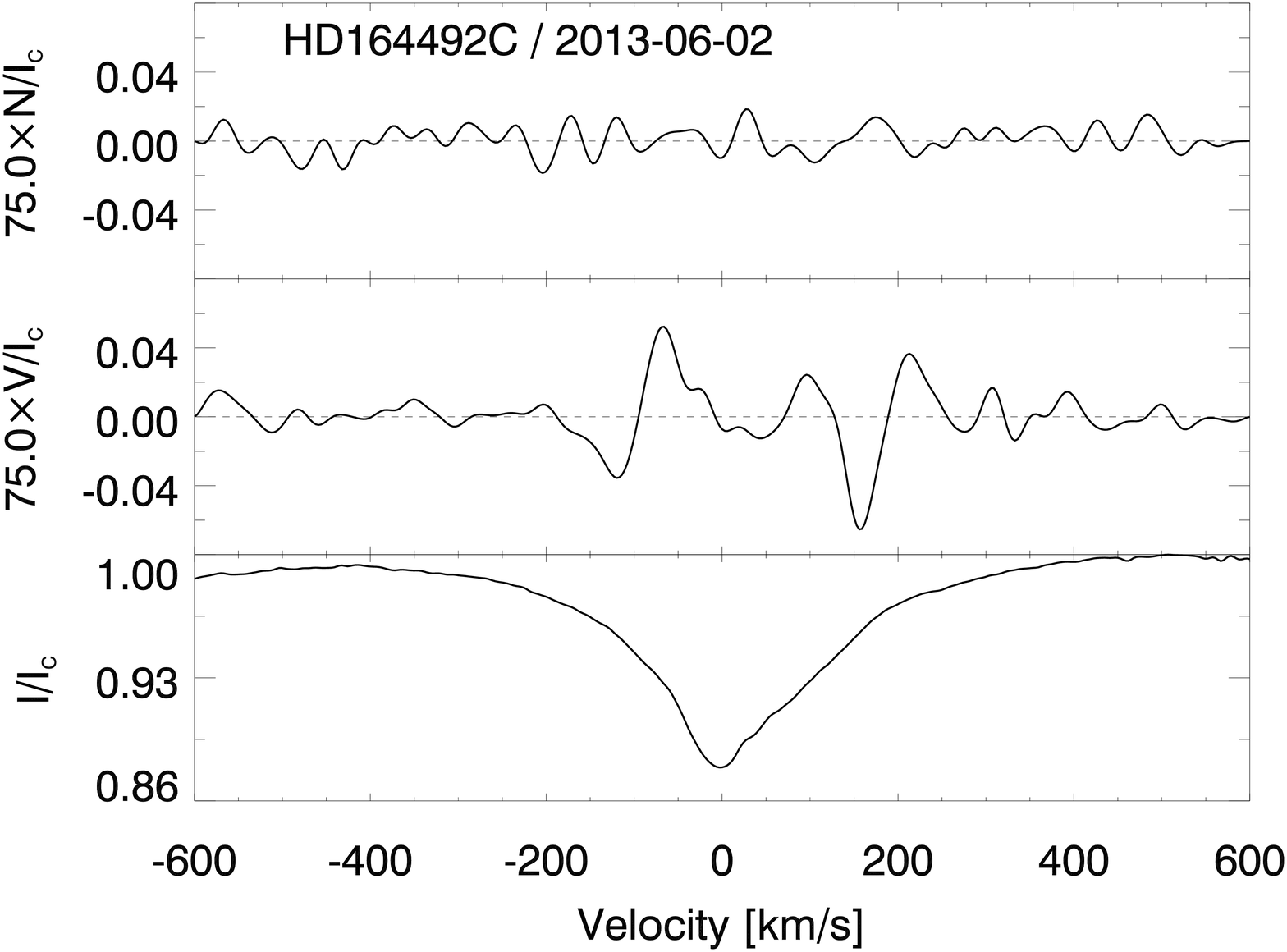}
\includegraphics[width=0.24\textwidth]{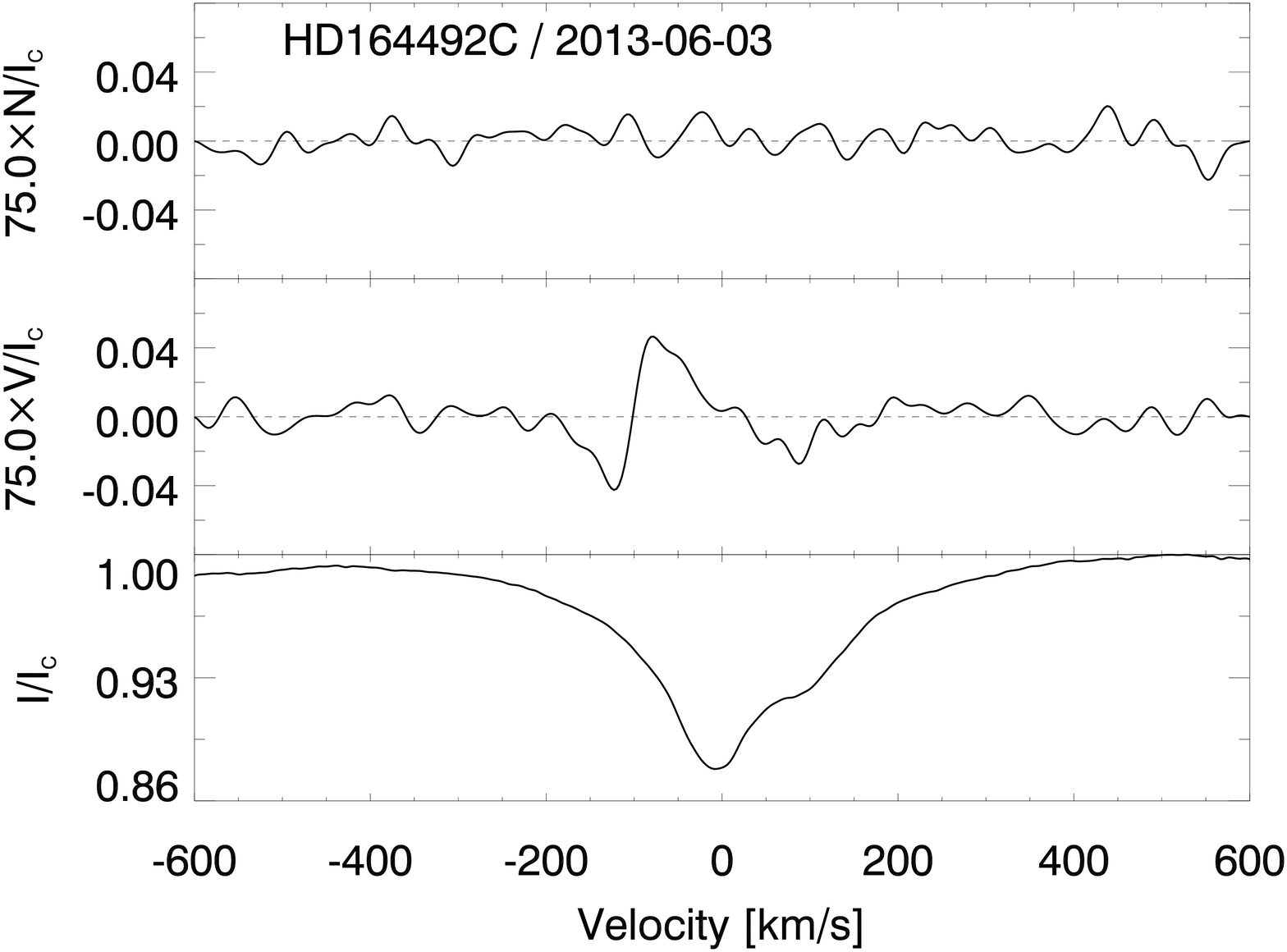}
\caption{
$I$, $V$, and $N$ SVD profiles  obtained for HD\,164492C for both nights.
The $V$ and $N$ profiles were expanded by 
a factor of 75 and shifted upwards for better visibility.
}
\label{fig:SVD}
\end{figure}

\begin{figure}
\centering
\includegraphics[width=0.24\textwidth]{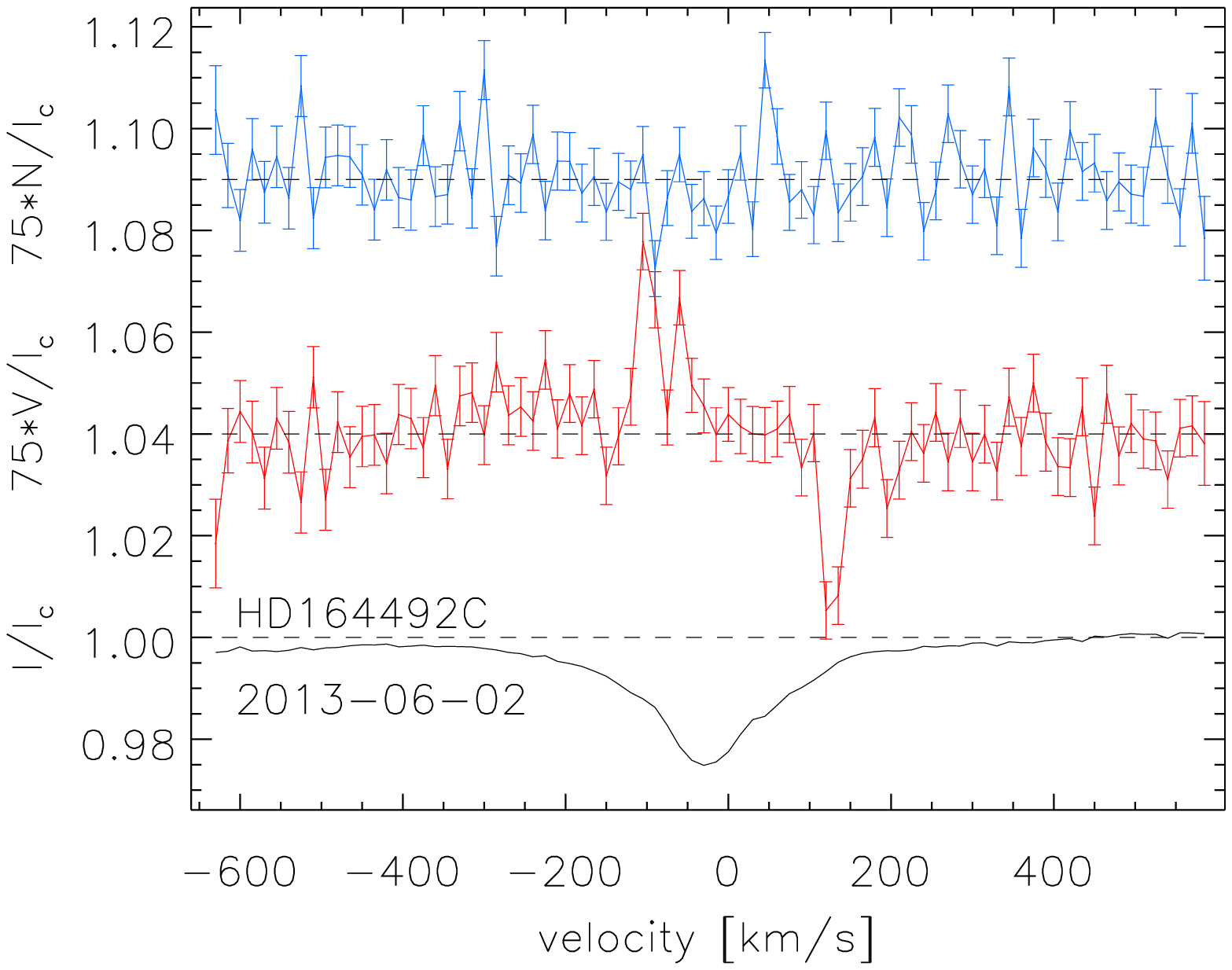}
\includegraphics[width=0.24\textwidth]{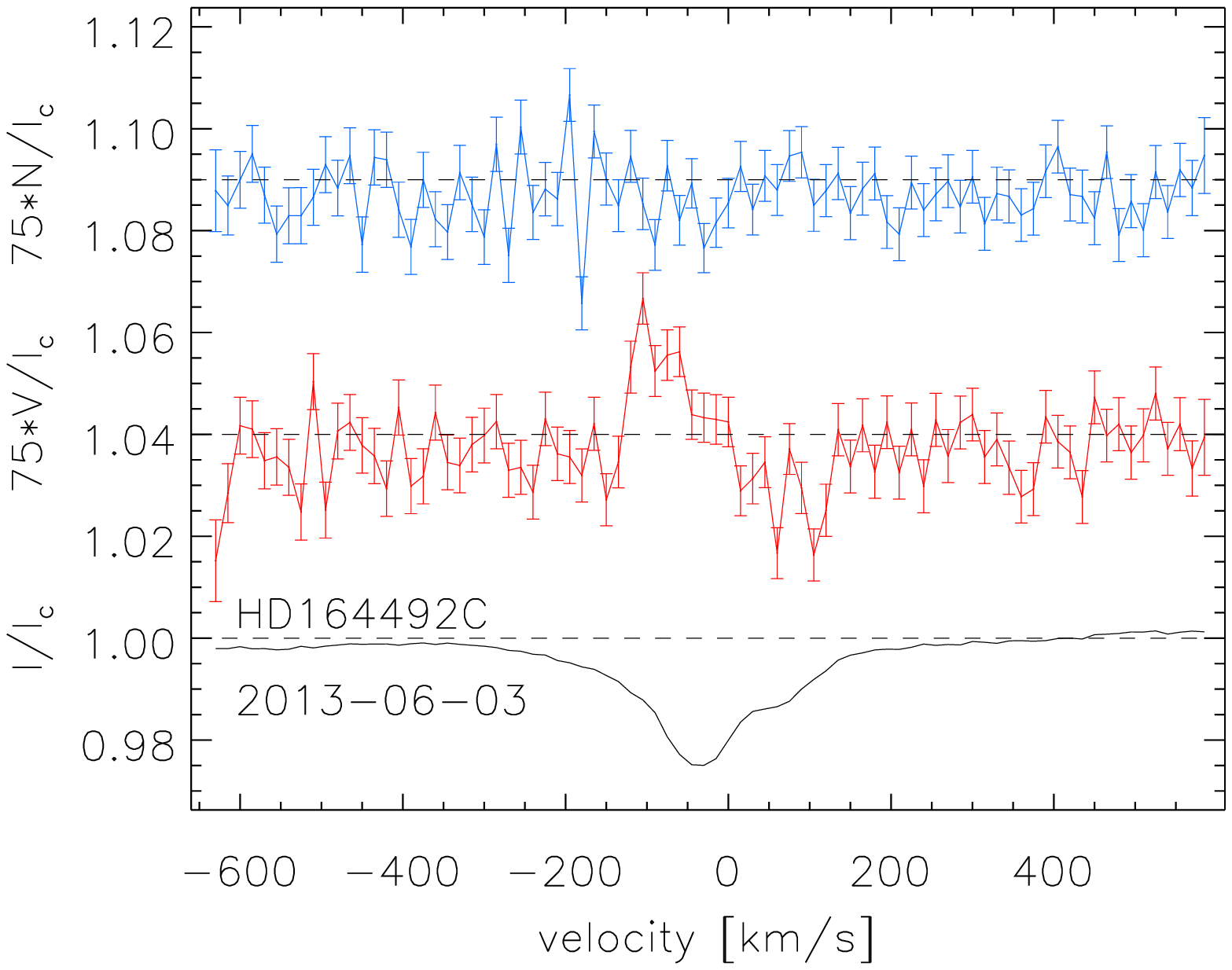}
\caption{
$I$, $V$, and $N$ LSD profiles (from bottom to top) obtained for HD\,164492C.
The $V$ and $N$ profiles were expanded by 
a factor of 75 and shifted upwards for better visibility.
}
\label{fig:lf}
\end{figure}

One software package used to study the magnetic field in HD\,164492C, the so-called  multi-line Singular Value Decomposition 
(SVD) method 
for Stokes Profile Reconstruction was recently introduced by Carroll et al.\ (\cite{carroll2012}). 
The results obtained with the SVD
method using about 80 lines in the line mask, including He lines and avoiding 
hydrogen and telluric lines, are presented in Fig.~\ref{fig:SVD}. The line mask was constructed using the VALD database 
(e.g.\ Kupka et al.\ \cite{kupka2000}).
Observations on both nights show definite detections with a false-alarm probability (FAP) lower than $10^{-10}$.
For the second software package we used the LSD technique 
(Donati et al.\ \cite{Donati1997}; Kochukhov et al.\ \cite{koch2010}).
The details of the analysis procedure
can be found in Makaganiuk et al.\ (\cite{Makaganiuk2011}) and Fossati et al.\ (\cite{Fossati2013}).
About 170 lines were used in the line mask,
but a test with the 80 lines used for the SVD program package did not show 
any significant difference in the results.
Fig.~\ref{fig:lf} shows that, similar to the SVD treatment, observations on both nights show definite detections 
with  FAPs lower than $10^{-10}$.  

\begin{figure}
\centering
\includegraphics[width=0.48\textwidth]{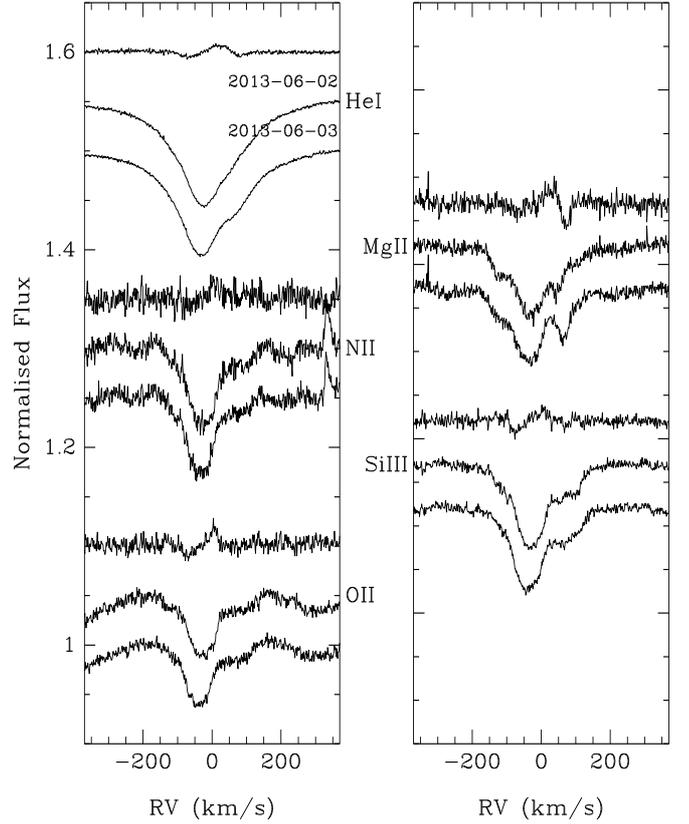}
\caption{
Average profiles (except for the \ion{Mg}{ii} 4481 line) 
calculated using several unblended lines: five \ion{Si}{iii} lines, ten \ion{O}{ii} lines, and four \ion{N}{ii} lines.
For comparison, we also present the average of ten \ion{He}{i} lines. 
The emission feature on the red side
of the \ion{N}{ii} profile belongs
to the \ion{O}{iii} 5006.8\,\AA{} line.
For each element, the upper spectrum presents the difference between
the average profiles from both nights. The apparent features in these difference spectra are mostly 
due to the variation of the radial velocity from one night to the next.
}
\label{fig:diff}
\end{figure}

\begin{figure}
\centering
\includegraphics[width=0.32\textwidth]{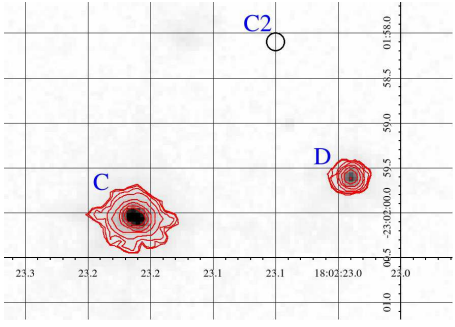}
\caption{
The HST WFPC2 image of HD\,164492C and its surroundings obtained during a short 30.8\,s exposure in the F502N filter.
North is up and east to the left.
The image size is $\sim 5''\times 3\farcs5$. 
Equatorial coordinates are shown, where the vertical axis is DEC and the horizontal axis is RA.
The contours are on logarithmic scale.
The letters C, D, and C2 refer to the notations of X-ray sources
as detected by Rho et al.\ (\cite{rho04}) using observations with the {\em Chandra} X-ray observatory. 
}
\label{fig:hst}
\end{figure}

The changes in the shape of the Stokes $I$ and $V$ profiles during two consecutive nights 
suggest that HD\,164492C is either a spectroscopically variable
star because of chemical spots or it is
a double-lined spectroscopic binary in which
the lines of the primary and the secondary appear strongly blended on both nights.
The presence of the weak \ion{He}{ii} 4686\,\AA{} line in the central strongest component indicates a spectral type B1
(for normal He content) or B1.5 (for enhanced He), implying
an effective temperature of about 24\,000--26\,000\,K (Nieva\ \cite{Nieva2013}). Moreover, the He lines 
appear in normal strength for the $T_{\rm eff}$ estimated from the
metal lines. Therefore a He-strong star can be excluded.
To better understand the origin of the variability of this system we investigated the behaviour of a 
few individual elements.
In Fig.~\ref{fig:diff}, we display the 
average profiles of the best clean lines identified in the HARPS spectra.
Remarkably, both the \ion{Si}{iii} and \ion{Mg}{ii} profiles exhibit several absorption peaks
in this plot.
In the same figure, the upper spectrum presents the difference between the average profiles from both nights
for each element.
The detected features in these difference spectra are mostly caused by the variation of the radial velocity 
from one night to the next, indicating that HD\,164492C cannot be considered as a single star.
According to the archive HST WFPC2 image of HD\,164492C  and its 
surroundings presented in Fig.~\ref{fig:hst}, 
the source C coinciding 
with HD\,164492C has an elongated shape, implying that we should see at least two stars in 
our HARPS spectra.
If we assume that we observe a binary system, a reasonable interpretation of the observed 
line profiles might be that the main absorption peak corresponds to one star with 
$v\,\sin\,i=55$\,km\,s$^{-1}$ that is superimposed on a broad-lined star with $v\,\sin\,i$ of 
about 140\,km\,s$^{-1}$. This second fast-rotating star would contribute to the left and right wings of the 
profiles presented in Fig.~\ref{fig:diff}.
It is possible that this star also has surface chemical patches, since 
profiles of different elements look somewhat different. 
On the other hand, the relatively rapid radial velocity variation in the spectrum of the primary
($\sim$4 km\,s$^{-1}$ per day) indicates that this star should have a close companion. Thus, we expect that the 
system HD\,164492C is at least a triple system.

\begin{figure}
\centering
\includegraphics[width=0.37\textwidth]{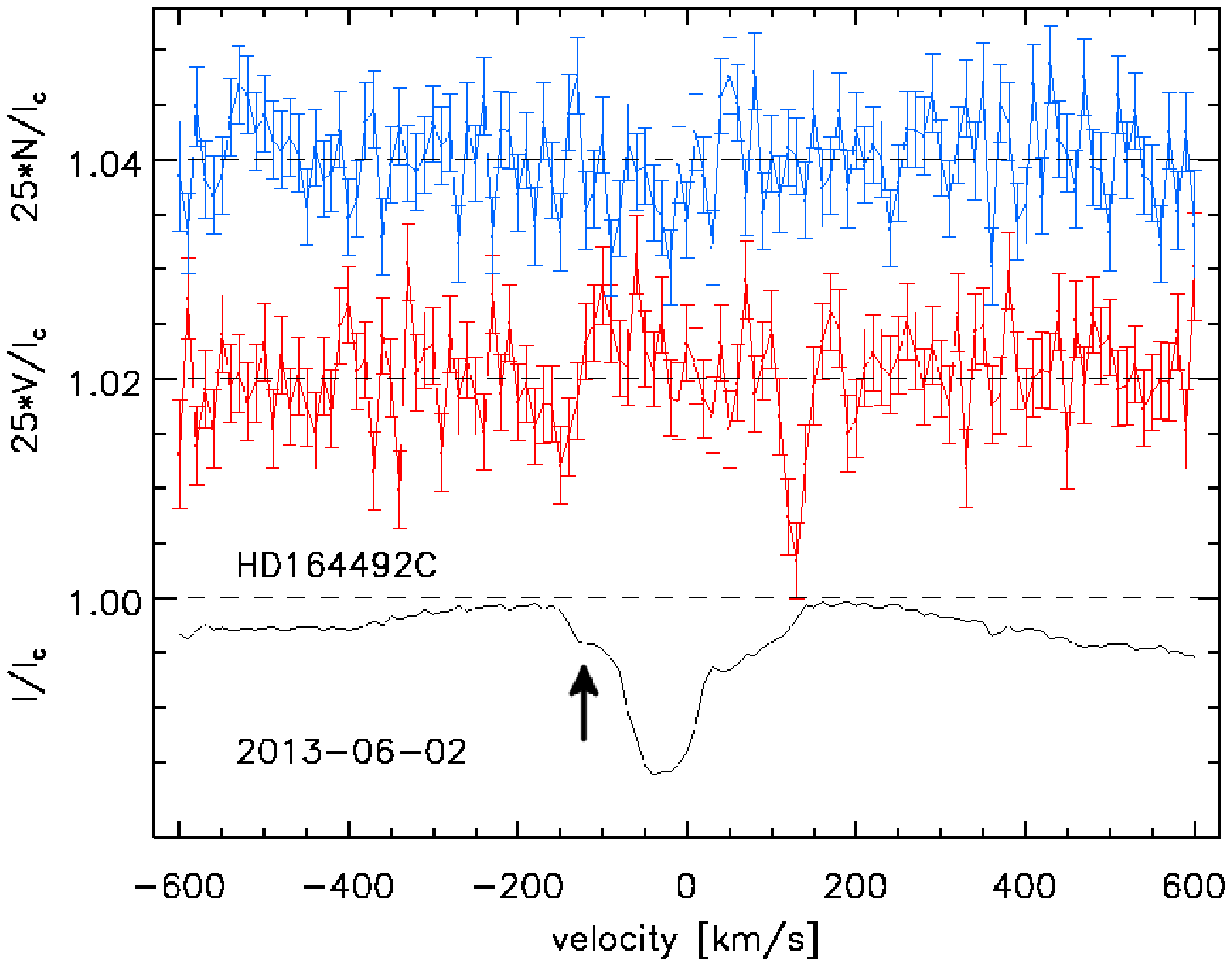}
\includegraphics[width=0.37\textwidth]{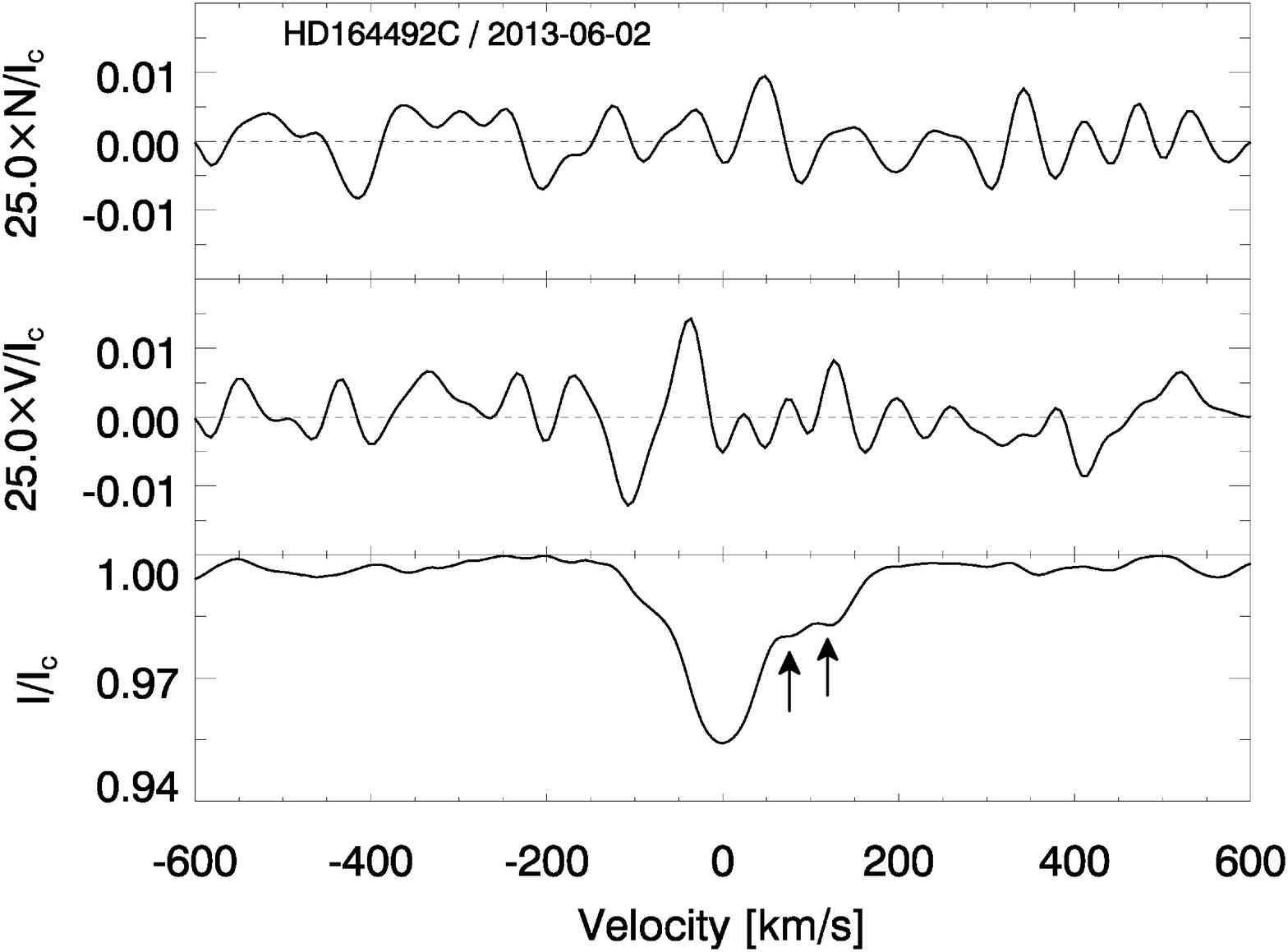}
\caption{
$I$, $V$, and $N$ profiles obtained for HD\,164492C during the first night.
{\it Top panel}: LSD profiles using metal lines.
{\it Bottom panel}: SVD profiles using \ion{Si}{iii} lines.
The $V$ and $N$ profiles were expanded by 
factors of 25 and shifted upwards for better visibility.
Additional absorption peaks are indicated by arrows. 
}
\label{fig:metal}
\end{figure}

In the top panel of Fig.~\ref{fig:metal},
we present $I$, $V$, and $N$ LSD profiles obtained
with the line mask excluding broad \ion{He}{i} lines.
In the bottom panel we show the SVD analysis where only 24 \ion{Si}{iii} lines are included.
In the SVD Stokes~$I$ \ion{Si}{iii} line profile, the blend in the red wing of the primary appears double-peaked,
while one more peak in the blue wing of the primary is best visible in the LSD Stokes~$I$ profile.
The measurement of the strength of the mean longitudinal magnetic 
field using the first-order moment method is described by Rees \& Semel (\cite{Rees79})
and requires measuring the equivalent width of Stokes $I$.
Because of the strong blending
of the components, we are unable to estimate the strength of the detected magnetic field.
Given the complex configuration and shape of the Stokes profiles, it is currently also impossible
to conclude exactly which components possess a magnetic field.
The detection of a significant Stokes $V$ signature in the SVD and LSD profiles at about $-$100\,km\,s$^{-1}$ and 
$+$150\,km\,s$^{-1}$ from the line core of the primary suggests that
more than one component might hold a magnetic field. 
The Stokes $V$ profile obtained on June~3 is well centred on the position of the primary.
Assuming that HD\,164492C is a single star, with the SVD and LSD methods we obtain
results very similar to those obtained 
with FORS\,2: between 500 and 700\,G for the first night, and 400 and 600\,G for the second night.

\section{Discussion}
\label{sect:disc}

Using FORS\,2 and HARPS in the framework of our ESO Large Program 191.D-0255, we detected a 
magnetic field in the poorly studied system HD\,164492C.
Although HD\,164492C appears to be a multiple system,
the multiplicity configuration is currently unclear and cannot be better elucidated without additional
observations.

X-ray emission from HD\,164492C is firmly 
detected using  {\em Chandra} observations, but is blended with a nearby unidentified X-ray source
(component C2; Rho et al.\ \cite{rho04}).
The total X-ray luminosity of these two marginally
spatially resolved sources is $2\times 10^{32}$\,erg\,s$^{-1}$, with
both components having similar X-ray brightness.
The component C2 shows X-ray variability and is harder in X-rays than HD\,164492C.  

To identify and model the physical processes that are responsible for the generation of magnetic fields
in massive stars, it is important to understand the formation mechanism of magnetic massive stars. 
Although the Trifid Nebula has often been 
studied, its distance is not accurately known.
Rho et al.\ (\cite{rho08}) reviewed literature values between 1.68 and 2.84\,kpc and adopted a
distance of about 1.7\,kpc. Cambr{\'e}sy et al.\ (\cite{cambresy11}) found
$2.7\pm0.5$\,kpc in their analysis of new near- and mid-infrared data.
Torii et al.\ (\cite{torii11}) used both 1.7 and 2.7\,kpc in their
discussion. Even shorter distances of 816\,pc and 1093\,pc
were estimated by Kharchenko et al.\ (\cite{kharchenko05,kharchenko13})
by combining proper-motion data with optical and near-infrared photometry
in their cluster analysis, respectively. 
The spatial distribution of the components of the multiple system ADS\,10991 and the photometric study by 
Kohoutek et al.\ (\cite{koho99}), which revealed almost the same E(B-V) values for components A--C, suggest that 
these components build a physical system in the nucleus of the Trifid nebula.
 
The age of the Trifid Nebula is only a few 0.1\,Myr
according to Cernicharo et al.\ (\cite{cernicharo98}),
who considered the spatial extent of the \ion{H}{ii} region.
The age of the cluster M20 and the time interval of the star formation in
this cluster, of which the system HD\,164492 is a member, can probably be larger, of the order of 1\,Myr.
Torii et al.\ (\cite{torii11})  argued that the formation of first-generation stars in the Trifid nebula, 
including the main ionising 
O7.5 star HD\,164492A, was triggered by the collision of two molecular clouds on a short time-scale of $\sim$1\,Myr. 
New insights into the understanding of massive star formation in the Trifid Nebula 
can be expected from a recently established large project based on infrared and X-ray observations 
of 20 massive star-forming regions, among them 
the Trifid Nebula (Feigelson et al.\ \cite{feig2013}). 

The presented first detection of a magnetic massive multiple system in one
of the youngest star-forming regions implies that this system
may play a pivotal role in our understanding of the origin of magnetic fields in massive stars. 
Future spectropolarimetric monitoring of this system is urgently needed to better characterise the components, their orbital 
parameters, and the magnetic field topology.

\begin{acknowledgements}
We thank J.~Ma\'iz Appell\'aniz, A.~de~Koter, A.~Herrero, and F.~Schneider for useful comments. 
T.M. acknowledges financial support from Belspo for contract PRODEX GAIA-DPAC. 
\end{acknowledgements}

\end{document}